\documentclass[%
 reprint,
 superscriptaddress,
 amsmath,amssymb,
 floatfix,
]{revtex4-2}

\usepackage{graphicx}
\usepackage{subfig}
\usepackage{floatrow}
\usepackage{dcolumn}
\usepackage{bm}
\usepackage[dvipsnames]{xcolor}
\usepackage{soul}
\usepackage{hyperref}
\usepackage{mathrsfs}
\usepackage{physics}
\usepackage[abs]{overpic}

\begin{document}

\title{Ionic blockade in a charged single-file water channel}

\author{Shusong Zhang}
\affiliation{
School of Physical Science and Technology, Northwestern Polytechnical University, Xi'an, 710072, China}

\author{Li Fu}
\affiliation{%
  Univ Lyon, Ecole Centrale de Lyon, CNRS, ENTPE, LTDS, UMR5513, 69130 Ecully, France
  }%
  
\author{Yanbo Xie}
\email{ybxie@nwpu.edu.cn}
\affiliation{
School of Physical Science and Technology, Northwestern Polytechnical University, Xi'an, 710072, China}
\affiliation{
School of Aeronautics and Institute of Extreme Mechanics, Northwestern Polytechnical University, Xi’an, 710072, China}

\date{\today}

\begin{abstract}

The classical continuum theories fail to describe the ionic transport in Angstrom channels, where conduction deviates from Ohm's law, as attributed to dehydration/self-energy barrier and dissociation of Bjerrum ion-pairs in previous work. Here we found that the cations are strongly bound to the surface charge that blockade the ionic transport in a single-file water channel, causing nonlinear current-voltage responses. The presence of free ions significantly increased the probability of bound ions being released, resulting in an ionic current. We found that ionic conduction gradually becomes Ohmic as surface charge density increases, but the conduction amplitude decreased due to increased friction from bound ions. We rationalized the ionic transport by 1D Kramers' escape theory framework, which well described nonlinear ionic current, and the impact of surface charge density on turning to Ohmic system. Our results possibly provide an alternative view of ionic blockade in Angstrom channels.

\end{abstract}

\maketitle

\section{\label{sec:introduction}Introduction}

Nanofluidics steps forward to the smallest scale. The recent studies showed the ionic transport within the Angstrom-scale channel is not capable of being described by the classical nanofluidic theories~\cite{berezhkovskii_single-file_2002,esfandiar_size_2017,mouterde_molecular_2019,robin_modeling_2021,kavokine_fluids_2021,li_breakdown_2023,kavokine_ionic_2019,kavokine_interaction_2022}, in particular when the Coulomb interaction of ions to the wall and ion-ion interactions are significantly reinforced in such small confinement~\cite{kavokine_ionic_2019,kavokine_interaction_2022}.
The contrast of dielectric properties between solution and surrounding substrate introduced a self-energy barrier for ions penetrating into the ion channels~\cite{parsegian_energy_1969,teber_translocation_2005, krems_ionic_2013, kaufman_coulomb_2015,kaufman_ionic_2017}, named as ionic Coulomb Blockade. The joint actions of self-energy barrier and ion-dehydration~\cite{zwolak_quantized_2009,li_anomalous_2021,yu_dehydration_2019} induced an energy barrier for ionic transport, producing a non-linear rising of conductance deviating from Ohm's law~\cite{feng_observation_2016}, which was considered as the principles of ion selectivity in ion channels in previous pioneer works~\cite{kaufman_coulomb_2015,kaufman_ionic_2017}. 
More recently, Kavokine et al.~\cite{kavokine_ionic_2019} achieved a groundbreaking theory framework in the ionic blockade which was dominated by the ion pairs in the channel studied by Brownian dynamics simulations, current only existing when the surface charge was fractional, known as fractional Wien effect.

However, the impact of surface charge density on the ionic conduction was still unknown, in particular the charge density increases as presence of more charged sites, where each bind-site was monovalently charged according to the charging mechanism at dielectric surfaces by chemical dissociation or physical adsorptions~\cite{behrens_charge_2001,stein_surface-charge-governed_2004}. 

In this work, we performed all-atom molecular dynamics (MD) simulations of a single-file water channel in a (10,0) zigzag carbon nanotube (CNT), to study the ionic transport within heterogeneously charged CNT and its impact of surface charge densities. We found an identical number of cations were strongly bound to the surface charges as we found previously~\cite{xie_liquid-solid_2020}, yielding a non-conductive water channel, named as the bound ion system. However, when a free cation is present in the CNT, it repels the bound ion and significantly increases the releasing probability of bound ion, thus reducing the threshold electric field of ionic transport, we named as the free-bound ion system which is the main system we studied. 

In periodically charged systems, we found the free cation knocked the bound ions one by one, fundamentally different from the previous theory by Zhang $et\ al.$~\cite{zhang_conductance_2005,zhang_ion_2006,kamenev_transport_2006} that assumed the bound ions synchronously dissociated from surface charges. We found the system gradually turned to be Ohmic as rising of surface charge densities, however the ion mobility and conduction decreased due to reduced slip lengths in CNT. We rationalized the ionic transport process via 1D self-propelled Kramers' escape problem, which well described the ionic current in our MD simulations. Our work may be useful for understanding and designing the Angstrom-scale water channel for the purpose of energies and mimicking biological ion channels.

\section{\label{sec:system}System and Methods}
We performed all-atom MD simulations to explore the ionic transport in a single-file water channel by LAMMPS package~\cite{thompson_lammps_2022}. We built a (10,0) zigzag CNT with radius $R=3.93$\,\r{A} and length $L = 204.48$\,\r{A}, with periodic boundary conditions in all directions. The CNT was cut off from a pre-equilibrium system with connected reservoirs under 298\,K and 1\,atm, shown as Fig. S1 in the Supplemental Material (SM)~\cite{sm}. The surface charge density is determined by the number of carbon atoms charged with $-e$.

We used the SPC/E force field~\cite{berendsen_missing_1987} for water, and the model of $\text{NaCl}$ by Koneshan~\cite{koneshan_solvent_1998}, considering water-carbon interactions from previous studies~\cite{werder_water-carbon_2003}. The temperature of the liquid was maintained at 298\,K by using a Nos\'{e}-Hoover thermostat applied to the $x$ and $y$ degrees of freedom. The CNT and water molecules were kept rigid in all simulations. Long-range Coulomb interactions were computed using the particle-particle particle-mesh (PPPM) algorithm. We found the water molecules assembled as single-file in our systems, consistent with previous studies~\cite{won_effect_2006,thomas_water_2009,gravelle_large_2014,gravelle_anomalous_2016}. More details about simulation systems can be found in Section SI in SM~\cite{sm}.

\section{\label{sec:result}Results}

\begin{figure}[b]
\minipage{0.47\textwidth}
\centering
\subfloat{%
\begin{overpic}[width = \textwidth]{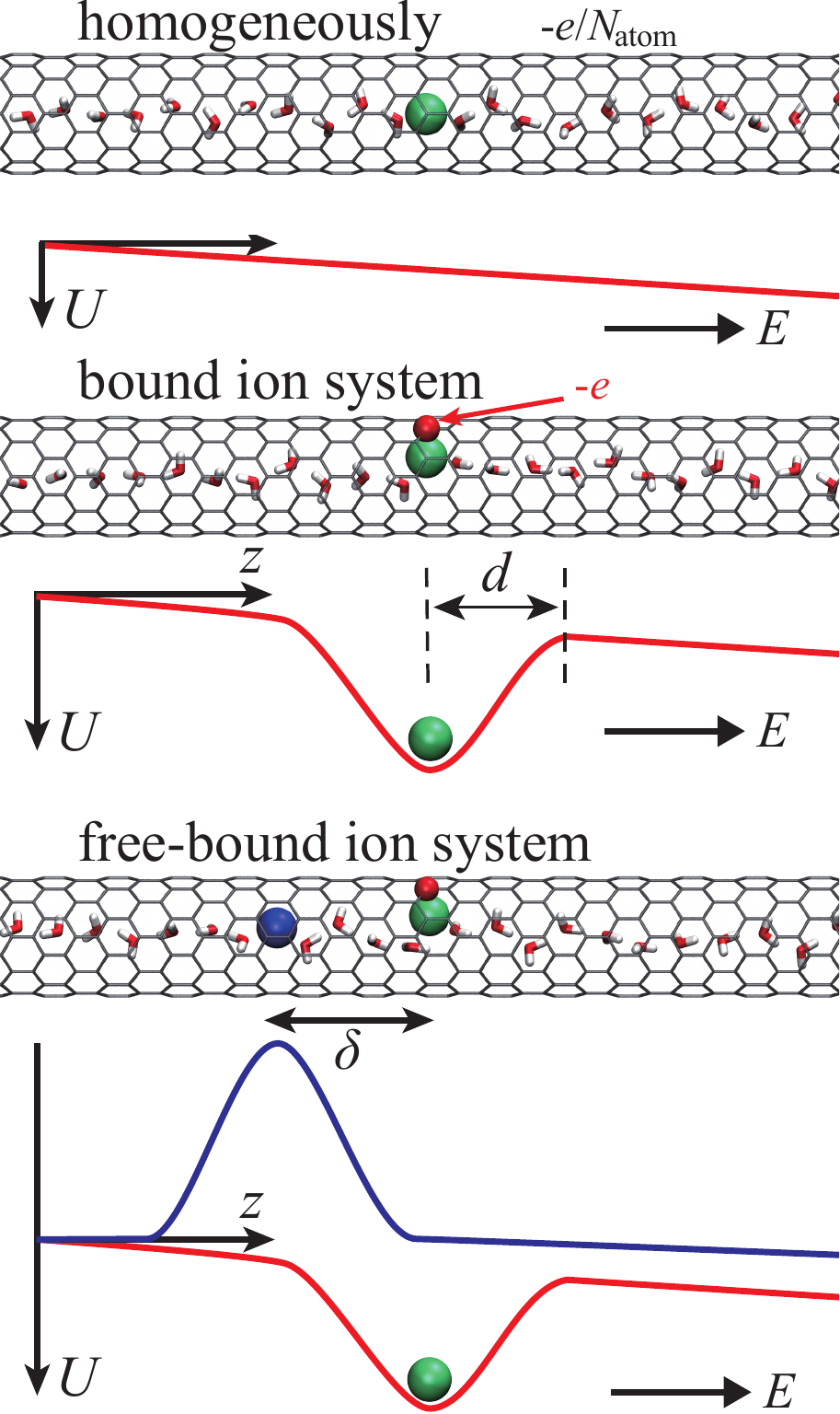}
\put(-1,190){(a)}
\put(-1,140){(b)}
\put(-1,77){(c)}

\end{overpic}
}
\endminipage
\hfill
\minipage{0.5\textwidth}
\centering
\subfloat{%
\begin{overpic}[width = \textwidth]{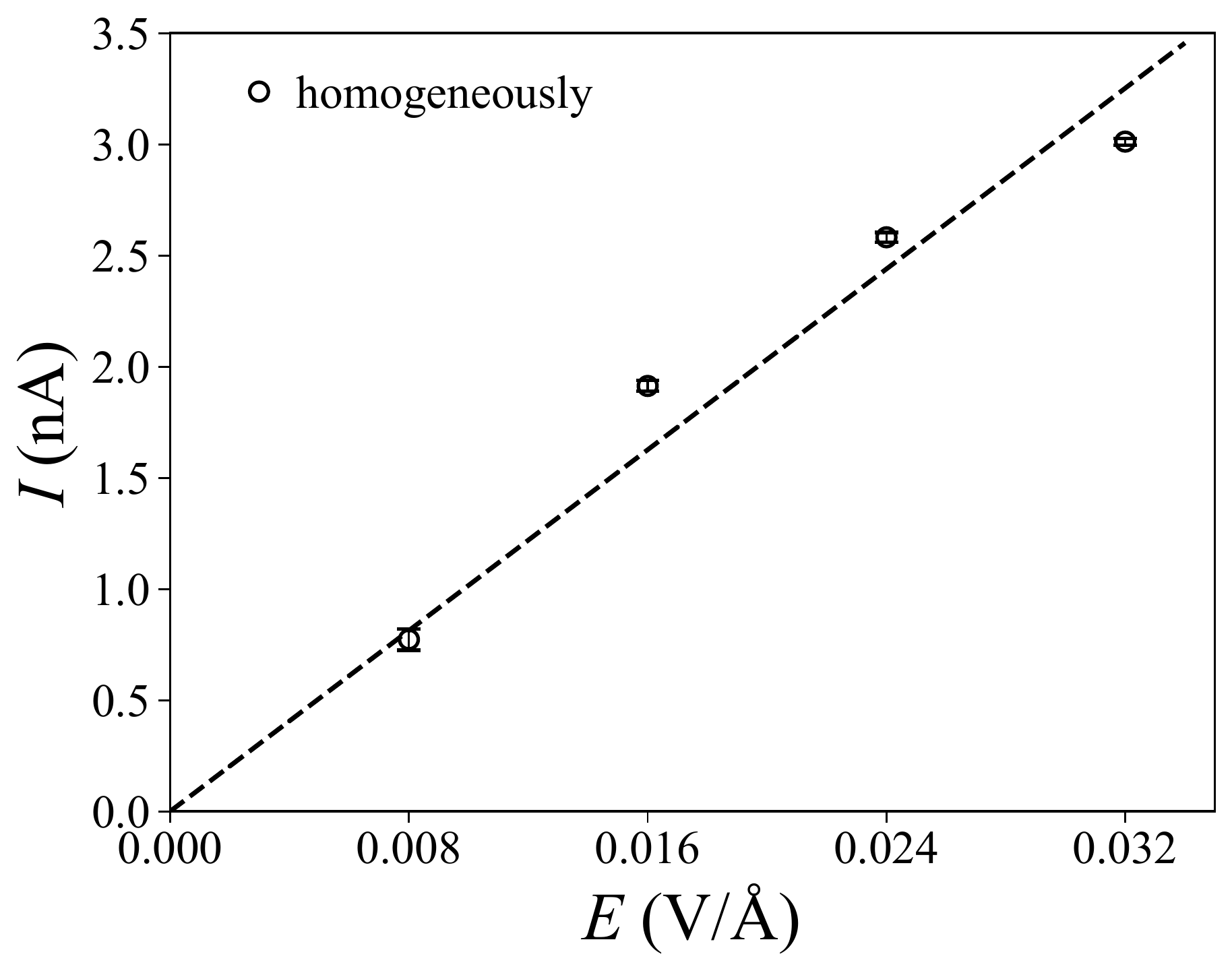}
\put(-7,90){(d)}
\end{overpic}
}
\hfill
\subfloat{%
\begin{overpic}[width = \textwidth]{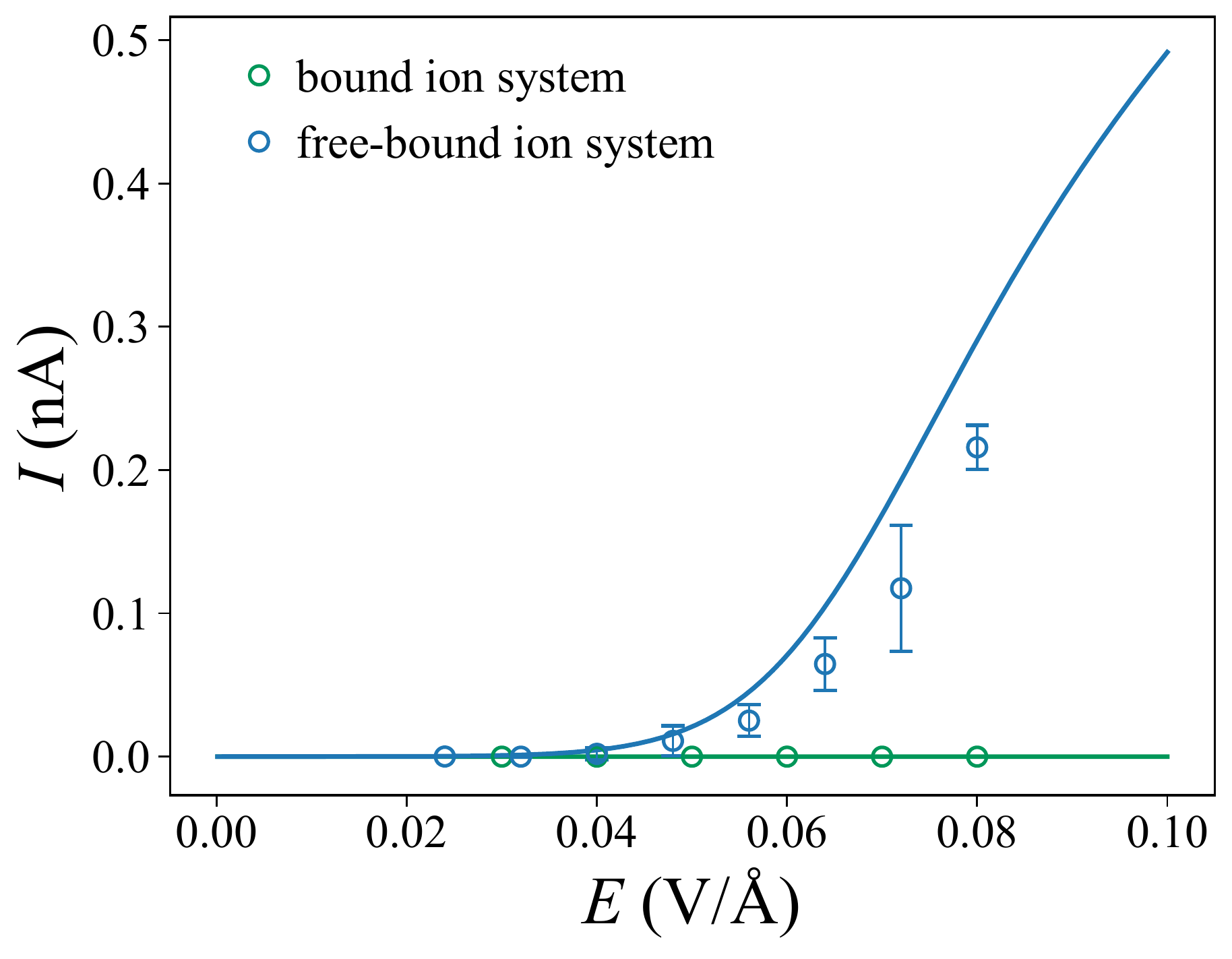}
\put(-8,90){(e)}
\end{overpic}
}

\endminipage

\caption{(a) is a snapshot for the homogeneously charged system where each single carbon atom is charged by $-e/N_\mathrm{atom}$ and $N_\mathrm{atom}$ is the number of carbon atoms of CNT. The red line represents the potential drop under an external electric field. (b) and (c) are snapshots of MD simulation for the bound ion system and free-bound ion system, respectively. The red line represents the potential well that originates from the surface charge (red particle) and external potential drop, trapping a cation (green particle). The blue line represents the Coulomb interaction of the free ion (blue particle). (d) Ionic current in the homogeneously charged CNT. The dashed line is a linear fitting of the $I-E$ curve. (e) Typical $I-E$ curves of bound ion system and free-bound ion system from simulations (dots) and the solid lines are theoretical results by Eq.~\ref{eq:I_single}. The error bars were derived from the standard deviations.}
\label{fig:1}
\end{figure}

We first studied homogeneously charged CNT where an elementary charge of $-e$ is homogeneously distributed on all carbon atoms ($i.e.$ each carbon atom is charged with $-e/N_\mathrm{atom}\sim -0.0005e$, where $N_\mathrm{atom}$ is the number of carbon atoms of CNT) and one $\text{Na}^+$ (green particle) is introduced in the CNT (see Fig.~\ref{fig:1}a). We applied an external electric field $E$ along $z$ direction, and computed the instantaneous ionic current by $\mathcal{I}(t)=\frac{e}{L} \sum_{i=1}^{n} v_{i}(t)$, where $n$ is the number of cations and $v_i (t)$ is the velocity of the $i$-th cation in $z$ direction. Every simulation was 
operated for 20\,ns to reach a steady state, and only the last 17\,ns were used for an average current $I$. We found that the current in the homogeneously charged CNT linearly increases with the electric field, represented as an Ohmic system shown in Fig.~\ref{fig:1}d. 

Then we focused on the ionic transport with a single carbon atom charged by $-e$ in length of periodic unit. As shown in Fig.~\ref{fig:1}b, in the bound ion system, a $\mathrm{Na^+}$ ion (green particle) was bound to the charged carbon atom (red particle) at CNT surface after thermal equilibrium, consistent with previous works~\cite{qiao_atypical_2003,li_effects_2019,xie_liquid-solid_2020}. Ionic current can only be formed when the bound ion escapes from the potential well induced by the surface charge in such single-file water channel. In contrast, in the free-bound ion system, the existing of a free $\mathrm{Na^+}$ ion (blue particle in Fig.~\ref{fig:1}c) provides an additional electrostatic repulsion to the bound ion. To keep the system neutral, we set $-e$ homogeneously distributed on all carbon atoms of CNT (an additional $-0.0005e$ per carbon atom).
The typical $I-E$ curves of bound ion system and free-bound ion system from simulations are shown as dots in Fig.~\ref{fig:1}e. 
Fig.~\ref{fig:1}e showed that both systems represented a non-conductive state at $E<0.04$\,V/\r{A}, as we found the bound ion was hardly released from the surface charge. Besides, the bound ion blocked the conductive path of the free ion passing through the single-file water channel in the free-bound ion system, exhibiting as a non-conduction state. 
As the rise of $E$, we observed an emerging of ionic current in the free-bound ion system above $\sim 0.04$\,V/\r{A}. Subsequently, the released bound ion drifts along the electric field, becoming a new free ion for the next knocking process. Reciprocally, the free ion which knocked the bound ion is then strongly bound by the surface charge. The ions repeated "knock-bound" cycles as a form of ionic current in such a system, with an example shown as Fig. S2 in SM~\cite{sm}. 
Fig.~\ref{fig:1}c illustrated the moment before knocking of the bound ion, which represents the distance between the free ion and bound ion $\delta$ that plays a role in the releasing probability for bound ion. 

Our MD results showed that the threshold $E$ of current generation in the free-bound ion system is significantly lower than the one in the bound ion system. Once the free ions existed in the CNT, which may penetrated from the reservoirs~\cite{li_breakdown_2023,zhou_field-induced_2023}, or be trapped between bound ions within current or volume flows, the mechanisms of ionic conduction is likely to be in a form of free-bound ion system, in the range of the electrical fields we studied. Thus, we will therefore focus on the free-bound ion system in later studies.

The classical mean-field theory is not capable to describe the ionic transport phenomena in the single-file channels. We rationalized the ionic conduction by Kramers' escape problems via 1D Fokker-Planck (FP) equation with probability density $W(z,t)$, as only escaping of the bound ions may form a current. To match the similar self-propelled escaping process in statistic mechanics~\cite{szamel_self-propelled_2014,geiseler_kramers_2016,woillez_activated_2019}, we introduced FP equation as,
\begin{equation} \label{eq:FP}
    \frac{\partial W}{\partial t}=D \left[\frac{\partial}{\partial z}\left(\frac{\partial (U(z)-F_\text{int}z) }{k_\text{B} T\partial z} W\right)+\frac{\partial^2  W}{\partial z^2} \right], 
\end{equation}

with potential 
\begin{equation} \label{eq:potential}
    U(z)=\phi(z) -eEz
\end{equation}

where $k_\text{B}$ is the Boltzmann constant, $T$ the temperature, $D$ the diffusion coefficient of ions related to mobility $\mu$ by Einstein relation $D=k_\mathrm{B} T \mu/e$.

The potential $\phi(z)$ considers the Coulomb interactions from all charged species, however only the free cation and surface charge play a role~\cite{teber_translocation_2005}, expressed as $\phi(z)=\phi_0(z)(1-e^{-L_n / \xi^\text{sc}})$. The $\phi_0(z)$ is the potential from the surface charge, and $L_n$ is the distance between two adjacent charged atoms(=204.48\,\r{A} for singly charged systems). The exponential decay term in $\phi(z)$ is from the electrostatic interaction of free ion that reduces the potential well, when the free ion is released from the adjacent surface charge with a distance of $L_n$. 
The $\xi^\text{sc}$ is the thermal length that we fitted from our MD simulation for a better consistency (see Fig. S3 in SM~\cite{sm}). With the potential drop of $-eEz$ due to external electric field, we have the system potential $U(z)$ of the bound ion.
The $F_\text{int}$ is the Coulomb repulsive force between the free ion and bound ion. If $F_\text{int}=0$ in Eq.~\ref{eq:FP}, we have the ordinary FP equation describing the escape problem of a particle from the potential well under an external electric field, i.e. bound ion system shown in Fig.~\ref{fig:1}b. 

We measured the potential distribution of surface charge $\phi_0(z)$ as well as ion-ion interactions $\phi_\text{int}(z)$ in simulation, which was numerically fitted in form of $\pm k_\text{B}T \frac{\xi}{x_T}e^{-|z|/ \xi}$~\cite{kavokine_ionic_2019}, with superscripts 'sc' and 'int' on $\xi$ and $x_T$ representing the potential from surface charge and ion-ion interaction respectively. The fitted $\xi^\text{sc}$ and $x_T^\text{sc}$ are slightly different from~\citet{kavokine_ionic_2019}, possibly due to the significant increase of Bjerrum length since the bound ion is exposed to vaccum-like region at CNT surface. Fitting details can be found in Fig. S3 and S4 in SM~\cite{sm}.

\begin{figure}[b]
\subfloat{%
\begin{overpic}[width=0.8\textwidth]{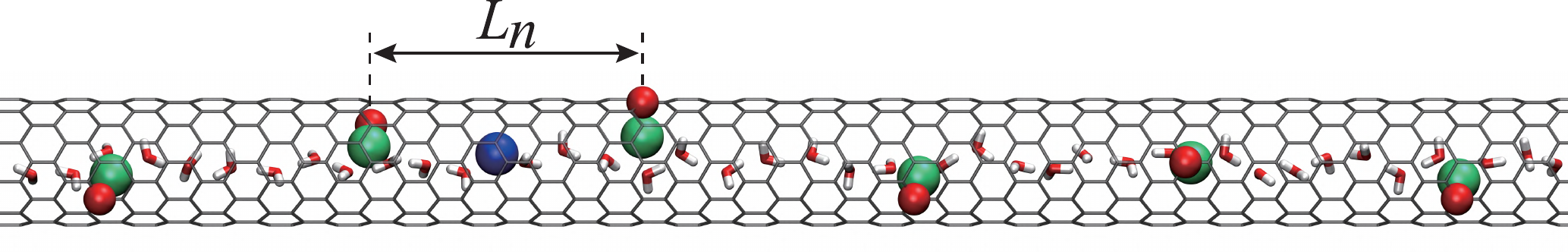}
\put(0,22){(a)}
\end{overpic}
}
\vspace{-0.5em}
\subfloat{%
\begin{overpic}[width=0.8\textwidth]{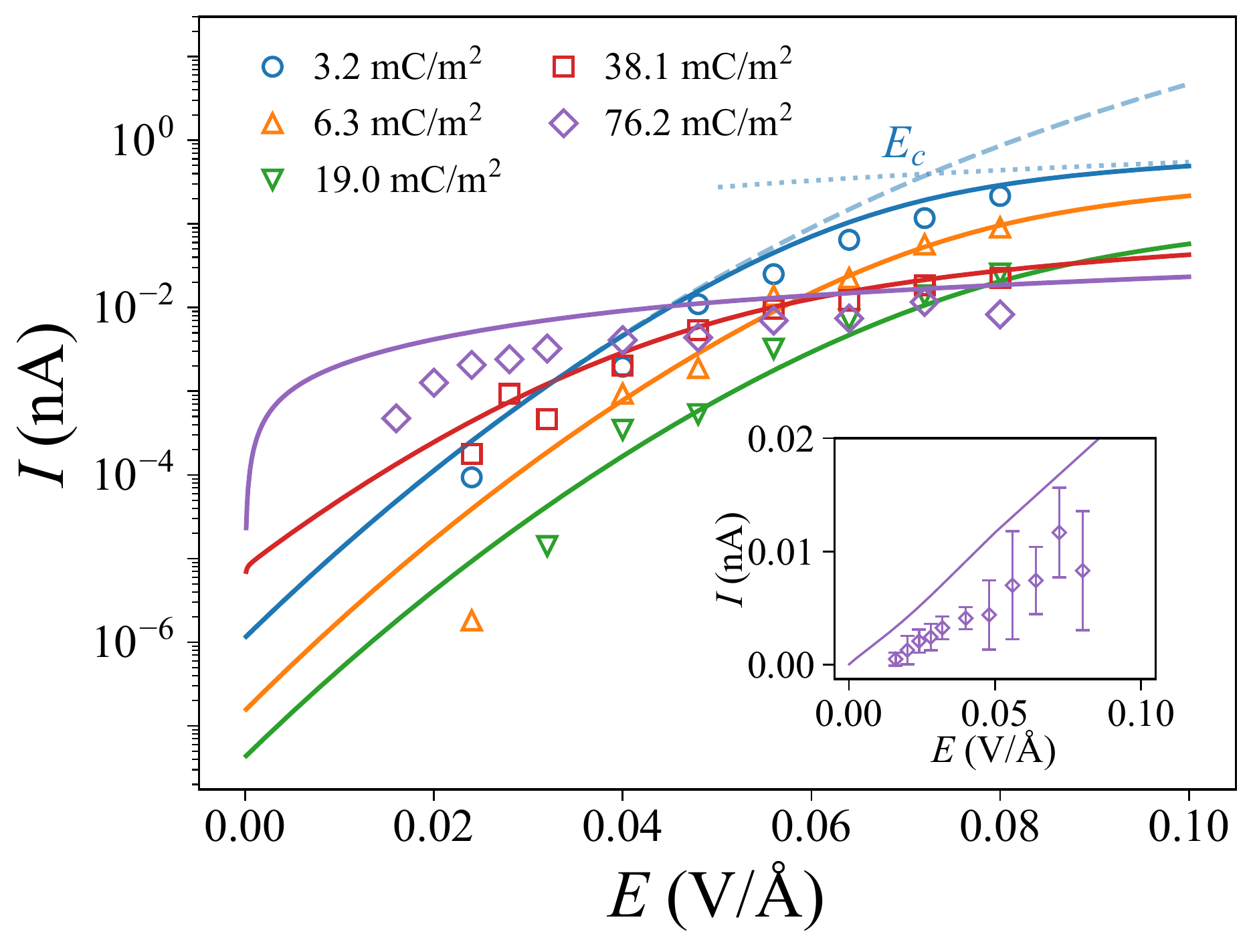}
\put(0,140){(b)}
\end{overpic}
}
\caption{(a) Snapshot of periodically charged CNT, where the red, green and blue particles are surface charge, bound ions, and free ion respectively. 
(b) $I-E$ curves at various surface charge densities. Dots represent the results from MD and solid lines are theoretical results using Eq.~\ref{eq:I_periodic} with mobility in Eq.~\ref{eq:mobility} and counted $\bar{\delta}$. We showed the Ohmic $I-E$ curve at $|\Sigma|=76.2$\,mC/m$^2$ as the inset figure. The error bars are derived by standard deviations. 
}
\label{fig:2}
\end{figure}

The motion of free ion brought electrostatic repulsive force and viscous friction force by electroosmosis to the bound ion, comprised of $F_\text{int}$. However, our MD simulation focused on regime where current was just emerged. As we found in our MD simulation, the free cation oscillated nearby the bound ion with a certain distance $\delta$ before the knock events (see Fig.~\ref{fig:1}c and stage \textcircled{2} in Fig. S2~\cite{sm}), with a negligible flow from electroosmosis. 
Furthermore, one can approximate $F_\text{int}$ as a constant before knocking the bound ion, expressed as $F_\text{int}(\bar{\delta}) = -\dd \phi_\text{int}(z)/\dd z|_{z=\bar{\delta}} = \frac{k_\text{B}T}{x_T^\text{int}} e^{-\bar{\delta}/\xi^\text{int}}$.
The $\delta$ slightly decreases with $E$ obtained from simulations in all surface charge densities $|\Sigma|$. For an approximation, here we use mean value $\bar{\delta}$ over the studied electric fields in each $|\Sigma|$. Thus, $\bar{\delta}$ is simply a function of $|\Sigma|$ and implemented in the theoretical models, shown as Fig. S5 in SM~\cite{sm}.

To solve the FP equation, we defined the surface charge locating at $z=0$ and introduced an absorbing boundary of the bound ion, where forces balanced by setting $\frac{\dd (U(z)-F_\text{int}(\bar{\delta})z)}{\dd z}|_{z=d'}=0$. We had the absorbing boundary as follows. 
\begin{equation} \label{eq:escape distance}
  d' = \xi^\text{sc} \ln \left( \frac{k_{\mathrm{B}} T (1-e^{-L_n / \xi^\text{sc}})}{(eE+F_\text{int}(\bar{\delta}))x_T^\text{sc}}\right). 
\end{equation}

We took the final absorbing boundary as $d = \max(d', 0)$ to avoid negative values of $d'$. Thus we could calculate the escape time for a bound ion from the potential well, under the joint actions of external electric field and a self-propulsion force~\cite{malakhov_time_1997, chupeau_optimizing_2020}, with details shown in SM~\cite{sm}:
\begin{equation}
  \tau_\text{es}=\frac{1}{D} \int_0^d e^{ \frac{-(U(z)-F_\text{int}z)}{k_\text{B}T}} \dd z \int_z^d  e^{ \frac{U(z')-F_\text{int}z'}{k_\text{B}T}} \dd z' . 
\label{eq:t_es}
\end{equation}

Furthermore, we could derive the current by counting the time of free drifting $\tau_\text{d}$ and time of escape $\tau_\text{es}$ for a single ion passing through a CNT with length $L$,
\begin{equation}
I = \frac{e}{\tau_\text{es}+\tau_\text{d}} . 
\label{eq:I_single}
\end{equation}

The $\tau_\text{d}$ can be calculated by $\tau_\text{d}=L/\mu E$, where $\mu$ is the electrophoretic mobility of ions, which are theoretically derived as a function of $|\Sigma|$ in the later texts. We calculated the current via Eq.~\ref{eq:I_single} shown as the solid lines in Fig.~\ref{fig:1}e for the bound ion system and free-bound ion system, and both matched well with the MD simulations. Our results showed the bound ion system didn't represent an obvious current, since the cation was not capable released from the surface charge (green line in Fig.~\ref{fig:1}e). However, once a free ion penetrated into the CNT, the probability of releasing the bound ion significantly increased resulting in current generation (blue line in Fig.~\ref{fig:1}e). 

Now, we turn to the impact of the surface charge densities on the ionic transport. We gradually increased the number of charged sites on the CNT surface while still keeping $L=204.48$\,\r{A} as a constant, to mimic the charging mechanisms at dielectric surfaces by dissociation of chemical groups or physical adsorption of charged species~\cite{behrens_charge_2001,stein_surface-charge-governed_2004}. 
The carbon atoms were charged by $-e$ and periodically distributed with a distance $L_n$ along $z$ direction, while randomly selected from the cross section of CNT . The total number of charged sites is $N=L/L_n$. At the equilibrium state, identical cations were bound to the charged sites at surface, with surface charge density of $\Sigma=-e/(2\pi R L_n)$.
Similarly, we introduced a free cation represented in Fig.~\ref{fig:2}a, with $-e$ homogeneously distributed on each carbon atom of CNT for a neutral system.

\begin{figure}[b]
\subfloat{%
\begin{overpic}[width=0.815\textwidth]{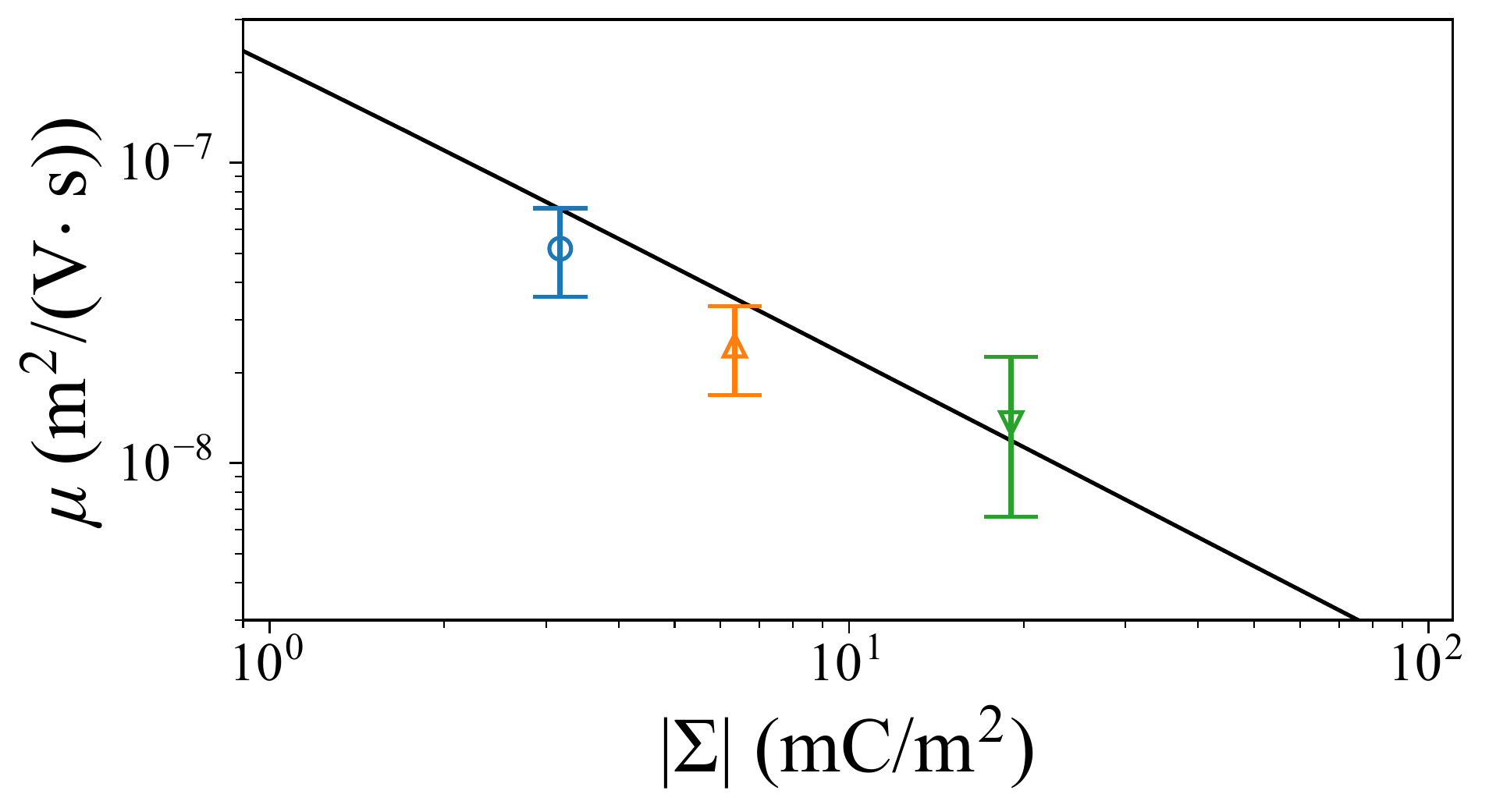}
\put(0,100){(a)}
\end{overpic}
}

\vspace{-1em}

\subfloat{%
\begin{overpic}[width=0.8\textwidth]{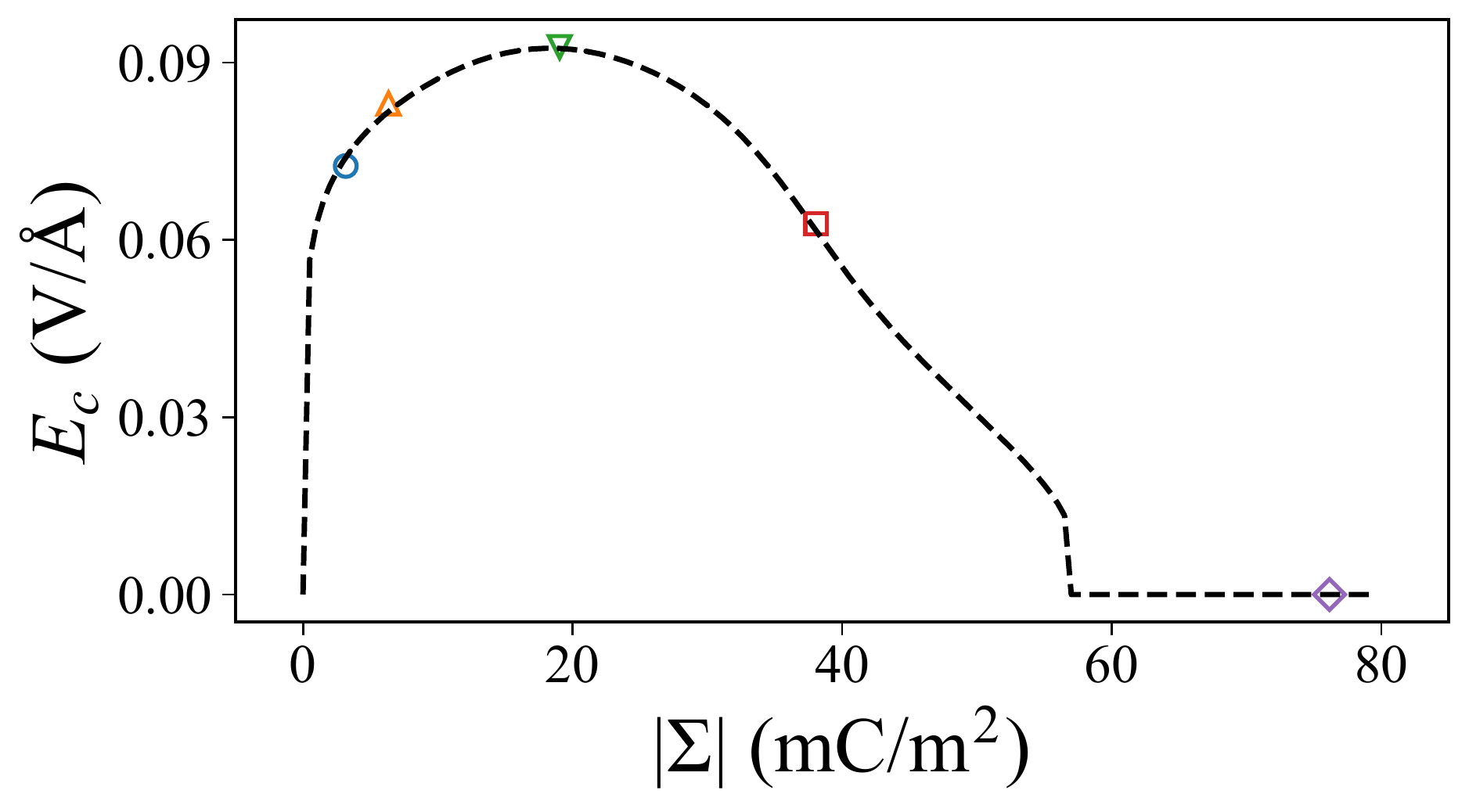}
\put(0,100){(b)}
\end{overpic}
}
\caption{
(a) Mobility of the free ion $\mu$ decreases as a function of surface charge density $|\Sigma|$. The dots are counted from MD simulation in free drifting processes and the solid line is calculated from Eq.~\ref{eq:mobility}. The error bars cover all statistic values.
(b) Threshold electric field $E_c$ of conduction state transition. The dots are calculated from $N\tau_\text{es}=\tau_\text{d}$, and the dashed line is an estimation by theories using fitted $\bar{\delta}- |\Sigma|$ and mobility from Eq.~\ref{eq:mobility}. }
\label{fig:3}
\end{figure}

The evolution of the current as a function of the electric field from simulations was shown as dots in Fig.~\ref{fig:2}b, with surface charge density $|\Sigma|$ increased from $3.2$ to $76.2$\,$\mathrm{mC/m^2}$, where the minimum $|\Sigma|$ corresponded to the system in Fig.~\ref{fig:1}c. The periodically charged systems also represented a non-conductive state at small $E$, then turning to conductive state as rising of $E$. However, there are two facts that we will focus in the following texts. The conductance decreases as $|\Sigma|$, and the change of threshold electric field for the conductive state transition as a function of $|\Sigma|$, which are illustrated in Fig.~\ref{fig:3}a and Fig.~\ref{fig:3}b, respectively.

We observed the free ion knocked the bound ion one by one from visualization of trajectories (see Video S1~\cite{sm}). To rationalize the current flux in periodically charged system, we assumed that all knocking processes of bound ions were identical. Hence, the escape time for a free ion transport through the length of CNT $L$ can be estimated as $N$ times of the individual knocking process $\tau_\text{es}$. The current can then be expressed as follows,
\begin{equation}
I=\frac{e}{N \tau_\text{es}+\tau_\text{d}}
\label{eq:I_periodic}
\end{equation}

One critical aspect for conduction is the ionic mobility or saying diffusion coefficient $D$ used in $\tau_\text{es}$ and $\tau_\text{d}=L/{\mu E}$. We introduced the slip length $b$ to characterise the effect of $|\Sigma|$ on mobility~\cite{huang_aqueous_2008,khair_influence_2009,joly_liquid_2006,xie_liquid-solid_2020}. Considering a free ion moves in a steady state in the CNT, the friction force originating from the wall and bound ions is balanced with the electrical driving force. More bound ions bring higher frictions, thus decreases slip length $b$ as well as the mobility of ions. We can thus derive $b$ as form of $b/b_0=1/(1 + 6\pi a b_0 k |\Sigma| /e)$ at heterogeneously charged surface following our previous work~\cite{xie_liquid-solid_2020}, where $b_0$ is slip length of a neutral CNT and $k=R/R_\mathrm{eff}$ is the correction factor due to the effective hydrodynamic radius $R_\mathrm{eff}$ of CNT. Finally, we derived the electrophoretic mobility $\mu$ via force balances of our system as a function of $|\Sigma|$,
\begin{equation}
\frac{\mu}{\mu_0}=\frac{3a}{R_\mathrm{eff} L}\frac{b_0}{1+6\pi a b_0  k|\Sigma|/e}
\label{eq:mobility}
\end{equation}

where $\mu_0=e/6 \pi \eta a$ is the mobility in the bulk fluids, known as Stokes-Einstein relation. The $\eta = 1.0$\,$\mathrm{mPa\cdot s}$ and $a=1.19$\,\r{A} are the viscosity of water and the effective hydrodynamic radius of cation, respectively. The $R_\mathrm{eff}$ can be estimated as $R_\mathrm{eff}=R-\frac{1}{2}\sigma_\text{CO}=2.3$\,\r{A}~\cite{fu_understanding_2018}. We obtained $b_0 = 676.5$\,nm from Green-Kubo method (Fig. S6 in SM~\cite{sm}). Finally, the theoretical value of $\mu$ is presented by the solid line in Fig.~\ref{fig:3}a. 

We derived the $\mu$ from simulations by linear fitting of the displacement of the free ion over time in free drifting process under certain $E$, which were plotted with open symbols in Fig.~\ref{fig:3}a. We found the theoretical prediction by Eq.~\ref{eq:mobility} matched well with the statistics from MD simulation. 
But the mobility in highly charged systems is difficult to be statistically counted as the short periodic length $L_n$ and hopping process of the bound ion after just knocked. As a consequence, the conduction significantly decreases with $|\Sigma|$ as seen in Fig.~\ref{fig:2}b due to the reduction of $\mu$ (or $b$). More details on the derivation of Eq.~\ref{eq:mobility} and statistics of mobility can be seen in Fig. S7 in SM~\cite{sm}. 
The diffusion coefficient $D = k_\mathrm{B} T b/(2\pi \eta R_\mathrm{eff} L )$ derived from our $\mu$ in our system is exactly the same as Detcheverry's work in a large slip limit~\cite{detcheverry_thermal_2012,detcheverry_thermal_2013}.

Then we statistically counted the $\bar{\delta}$, which shows the mean value $\bar{\delta}$ slightly decreases as rising of $|\Sigma|$ (see Fig. S5 in SM~\cite{sm}), increasing the repulsion force $F_\text{int}$ and probability of bound ion being knocked. As $L_n$ decreases, less water molecules need be squeezed out between the bound ion and free ion, resulting in a decrease in $\bar{\delta}$. Using Eq.~\ref{eq:I_periodic} combined with theoretical mobility Eq.~\ref{eq:mobility} and counted $\bar{\delta}$, we can calculate the theoretical current responses shown as solid lines in Fig.~\ref{fig:2}b.

Finally, to quantitatively evaluate the threshold electric field $E_c$ from the closed state to the open state of ionic transport, we employ the intersection between two extreme conditions: Arrhenius-type conduction $I_\text{es}=e/(N \tau_\text{es})$ where escape time is dominant shown as dashed line in Fig.~\ref{fig:2}b, and Ohmic-type conduction $I_\text{d} = e/ \tau_\text{d}$ where time of free drifting is dominant shown as dotted line in Fig.~\ref{fig:2}b. Here we define the electric field strength of intersections as the threshold electric field $E_c$ for conductive state transition, plotted as dots in Fig.~\ref{fig:3}b.

Meanwhile, we theoretically estimated $E_c$ by setting $N \tau_\text{es}=\tau_\text{d}$ using Eq.~\ref{eq:mobility} and fitted $\bar{\delta}$ (See Fig. S5b in SM~\cite{sm}), and exhibited as dashed line in Fig.~\ref{fig:3}b. The changes of $\mu$ do not affect $E_c$ since $\mu$ or $D$ is both included in the denominator of equations of $\tau_\text{es}$ and $\tau_\text{d}$.
The $E_c$ first rapidly increased to a certain value as the emergence of a bound ion in CNT, then reached to a saturated value in sparsely charged CNT when $L_n \gg \xi^\text{sc}$.
When $|\Sigma|$ reached $\sim$20\,$\mathrm{mC/m^2}$, $E_c$ gradually decreased as $|\Sigma|$ and finally vanished over 57\,$\mathrm{mC/m^2}$ that the systems turn to an Ohmic system. The reduction of $E_c$ was caused by the Coulomb interaction between free ion and bound ion that lowered the surface potential $\phi (z)$ as $|\Sigma|$ increases. The sudden vanishing of $E_c$ at 57\,$\mathrm{mC/m^2}$ is caused by that the lines of $I_\text{es}$ and $I_\text{d}$ are just tangent to each other.
When $I_\text{es}>I_\text{d}$ at above 57\,$\mathrm{mC/m^2}$, we could not find an intersection of $I_\text{es}$ and $I_\text{d}$, it comes to the Ohmic conduction regime. As a consequence, we have $E_c=0$.

\section{\label{sec:conclusion}Conclusion}

In this work, we studied the ionic transport in a single-file CNT by all-atom MD simulation and 1D self-propelled Kramers' escape problem. 
We found identical cations were strongly bound to the surface charge due to the reinforced Coulomb interactions in such small confinement, resulting in the blockade of ionic transport in single-file CNT. However, with an additional free cation in the CNT, the single-file channel becomes conductive under a certain electric field via knocking of bound ion, as we named the free-bound ion system. 
Then we studied the impact of surface charge densities by increasing the number of charged sites of CNT. We found a free ion knocked the bound ions one by one and repeated a "knock-bound" cycle thus forming an ionic current, whose mechanism is different from both of previous predictions by Zhang $et\ al.$~\cite{zhang_conductance_2005,zhang_ion_2006,kamenev_transport_2006} and the classical mean-field theories in nanofluidics. 
Thus, we used 1D self-propelled Kramers' escape problem to describe the ionic current as a function of electric field, which showed the current-voltage responses gradually turned into Ohmic system due to ion-ion interactions as surface charge densities rise.
We defined $E_c$ to quantitatively describe the state transition, by intersection of Arrhenius-type current dominated by escape time and Ohmic current dominated by free drifting. We found $E_c$ first increased to a saturated value in sparsely charged CNT, then decreased and finally vanished at $|\Sigma|\sim$57\,$\mathrm{mC/m^2}$.
In addition, we found a significant decrease of conductance as $|\Sigma|$ rises, which was attributed to the reduced ion mobility correlated to the slip length, as more bound ion brought stronger friction. The analytical equation well predicted the decrease of mobility in simulations as $|\Sigma|$ increases. 
Our work proposed a new mechanism of ionic blockade caused by the bound ion at surface, besides the Coulomb blockade and fractional Wien effects, which will possibly be helpful for the understanding of Angstrom-scale transport and applications of ion separations or energies. 

\begin{acknowledgments}
The authors thank Prof. Laurent Joly in Lyon-1 University for the insightful discussions. This work is supported by NSFC (NO.12075191, 12241201). 
\end{acknowledgments}

\bibliography{main}

\end{document}